\documentclass[9pt,conference]{IEEEtran}
\IEEEoverridecommandlockouts

\usepackage{amsmath, bbm}
\usepackage{amssymb}
\usepackage{amsfonts}
\usepackage{amsthm}
\usepackage{nicefrac}
\usepackage{bm}
\usepackage{hyperref}
\usepackage{cite}
\usepackage[utf8]{inputenc} 
\usepackage[T1]{fontenc}
\usepackage{graphicx}
\usepackage[font=small]{caption}
\usepackage{tikz}
\usetikzlibrary{positioning}
\usepackage{pgfplots}
\usepackage{epstopdf}
\usepackage[labelformat=simple]{subcaption}
\usepackage{multirow}
\usepackage{lipsum}
\usepackage{mathtools}

\usepackage{algorithm}
\usepackage[noend]{algpseudocode}
\algrenewcommand\algorithmiccomment[1]{%
  \hfill{\color{black!55}\itshape\(\triangleright\)\ #1}%
}

\long\def\comment#1{}

\newtheorem{Problem}{Problem}

\input{mysymbol.sty}

\title{\huge A Differentiable Digital Twin of Distributed Link Scheduling for Contention-Aware Networking
\thanks{
    Research was sponsored by the DEVCOM ARL Army Research Office and was accomplished under Cooperative Agreement Number W911NF-24-2-0008. 
    The views and conclusions contained in this document are those of the authors and should not be interpreted as representing the official policies, either expressed or implied, of the DEVCOM ARL Army Research Office or the U.S. Government. 
    The U.S. Government is authorized to reproduce and distribute reprints for Government purposes notwithstanding any copyright notation herein.
    \newline
    $^\ast$ These authors contributed equally to this work
    \newline
    Emails: $^\star$\{zhongyuan.zhao, yujun.ming, segarra\}@rice.edu, $^\ddag$\{kevin.s.chan, ananthram.swami\}.civ@army.mil}
}

\author{
Zhongyuan Zhao$^{\ast\star}$, Yujun Ming$^{\ast\star}$, 
Kevin Chan$^{\ddag}$, 
{Ananthram Swami}$^{\ddag}$, 
and {Santiago Segarra}$^{\star}$\\
\textit{$^\star$Rice University, USA \hspace{10mm}  \hspace{2mm}  $^\ddag$US Army DEVCOM Army Research Laboratory, USA}
}

\begin{document}
% \ninept
\renewcommand{\baselinestretch}{0.98}
\maketitle
\begin{abstract}
Many routing and flow optimization problems in wired networks can be solved efficiently using minimum cost flow formulations. 
However, this approach does not extend to wireless multi-hop networks, where the assumptions of fixed link capacity and linear cost structure collapse due to contention for shared spectrum resources. 
The key challenge is that the long-term capacity of a wireless link becomes a non-linear function of its network context, including network topology, link quality, and the traffic assigned to neighboring links. 
In this work, we pursue a new direction of modeling wireless network under randomized medium access control by developing an analytical network digital twin (NDT) that predicts link duty cycles from network context. 
We generalize randomized contention as finding a Maximal Independent Set (MIS) on the conflict graph using weighted Luby's algorithm, derive an analytical model of link duty cycles, and introduce an iterative procedure that resolves the circular dependency among duty cycle, link capacity, and contention probability. 
Our numerical experiments show that the proposed NDT accurately predicts link duty cycles and congestion patterns with up to a 5000× speedup over packet-level simulation, and enables us to optimize link scheduling using gradient descent for reduced congestion and radio footprint.
\end{abstract}
%

% \vspace{-0.05in}
\section{Introduction}\label{sec:intro}
A growing class of next-generation applications increasingly relies on self-organizing wireless multi-hop networks for flexible and infrastructure-light connectivity, spanning mobile ad-hoc networks, xG (device-to-device, wireless backhaul, and non-terrestrial coverage), vehicle-to-everything (V2X) and aerial networks, Internet-of-Things (IoT) and Machine-to-Machine (M2M) communications~\cite{kott2016internet,akyildiz20206g,chen2021massive,noor20226g,Tezergil2022,Xiao2024space}.
Distributed link scheduling~\cite{ni2012qcsma,joo2012local,zhao2022twc} is a key enabler of  self-organization, allowing transceivers to access shared medium without reliance on a central controller. 
Most distributed scheduling mechanisms fall into two categories: deterministic and randomized contention.
Deterministic approaches, exemplified by Backpressure routing and MaxWeight scheduling~\cite{neely2005dynamic,zhao2024tmlcn,zhao2023icassp,zhao2025icassp,joo2012local,zhao2022twc}, offer optimal throughput but often suffer from poor fairness and impose higher requirements on synchronization and coordination. 
In comparison, randomized contention, such as listen-before-talk protocols~\cite{Song2016coexistence} and carrier sensing multiple access (CSMA)~\cite{ni2012qcsma,geraci2025wi}, features with lower spectrum efficiency~\cite{Jindal13} but better fairness and critically, minimum coordination requirements.
As a result, randomized contention is widely used in wireless systems with heterogeneous devices and in shared or lightly licensed bands (e.g., ISM, mmWave)~\cite{Song2016coexistence,zhao2019tvt,geraci2025wi,Tezergil2022}.

The benefits of randomized contention in medium access control (MAC) come at the cost of significant challenges in networking.
Tasks like packet routing, computation offloading, and load balancing can be solved efficiently in wired networks using minimum-cost flow formulations~\cite{chen2025maximum}. 
Yet these approaches do not extend to wireless multi-hop networks, where the assumptions of fixed link capacity and linear cost structure collapse under contention for the shared medium, instead, they become functions of the network context, including the qualities of and the traffic assigned to a link and its neighbors.
The Backpressure-MaxWeight family of algorithms~\cite{neely2005dynamic,zhao2023icassp,zhao2024tmlcn,zhao2025icassp,joo2012local,zhao2022twc} addresses this challenge under deterministic scheduling by solving a Maximum Weighted Independent Set problem on the conflict graph each time slot, yielding conflict-free routing and scheduling decisions driven by congestion and distance gradient.
These benefits, however, diminish under randomized contention.
\begin{figure}
    \centering
    \includegraphics[width=0.95\columnwidth]{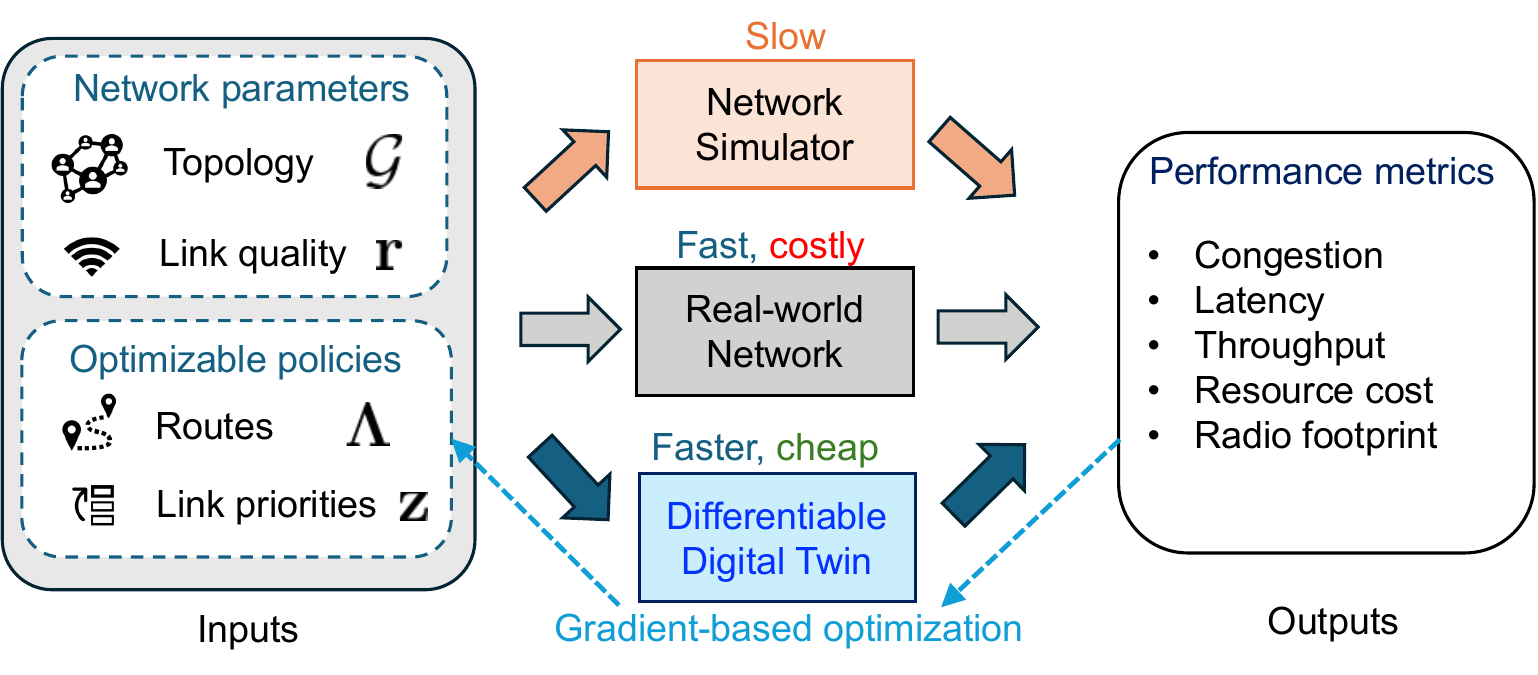}
     \vspace{-0.1in}
    \caption{Methods of wireless network analysis and optimization.
    } 
    \label{fig:model:dt}
     \vspace{-0.1in}
\end{figure}

Existing approaches to wireless networking under randomized contention typically rely on three strategies to compensate for the lack of reliable models for long-term wireless link capacity.
1) \textit{Ignore}: treat the wireless network as if it were wired by assuming link capacity to be a fixed fraction of the expected transmission data rate~\cite{Erfaniantaghvayi2025mobihoc}, or imposing a prescribed TDMA schedule~\cite{Tezergil2022}. 
2) \textit{Online probing}: infer link capacities on-demand using probing packets, assuming that new traffic requests arrive incrementally and do not significantly perturb the existing network state~\cite{chen2005ad,Pakzad2015}. 
3) \textit{Machine learning (ML)}: use data-driven models to predictively select routes or next hops based on historical observations~\cite{Ding2022tvt,kong2024covert,zhao2024congestionaware}.
These strategies, however, suffer from limitations such as unexpected congestion due to inaccurate capacity estimates, inefficient resource utilization, excessive probing or training overhead, and poor generalization across network topologies, traffic patterns, and MAC settings.
Moreover, the absence of a long-term capacity model forces network analysis and optimization to heavily rely on packet-level simulation~\cite{riley2010ns}, which is slow and scales poorly with network size.

In this work, we close this gap by developing an analytical network digital twin (NDT) that predicts link duty cycles~\cite{zhao2024tmlcn,zhao2023icassp,zhao2024congestionaware} under randomized MAC based on network context, including topology, link quality, and routing and scheduling policies. 
Such predictions 
can be directly translated into long-term link capacities as well as a variety of cost and performance metrics.
Compared to trainable alternatives~\cite{zhao2024tmlcn,zhao2023icassp,zhao2024congestionaware,li2023milcom}, our analytical NDT is lightweight and generalizes across diverse network topologies, traffic patterns, and MAC configurations.
As illustrated in Fig.~\ref{fig:model:dt}, the NDT enables rapid performance assessment without costly simulation, probing, or trial-and-error, offering a new methodology for wireless network analysis and optimization. 
% \noindent
% {\bf Contributions:} 
Our \textbf{contributions} are three-fold:
\begin{itemize}
    \item We generalize randomized contention in wireless multi-hop networks into a weighted variant of Luby's algorithm~\cite{luby1985simple} for solving a MIS problem on the conflict graph of the network.
    \item We develop an analytical NDT by deriving an close-form expression of link duty cycles under weighted Luby, and introducing an iterative procedure that resolves the circular dependency among duty cycle, link capacity, and contention probability. 
    % Unlike fixed capacity assumptions, this model predicts link duty cycle as a function of network states and parameters, enabling accurate congestion and performance prediction.
    \item Through numerical experiments, we demonstrate the accuracy and significant runtime speedup of our NDT relative to packet-level simulations, 
    and we apply it to gradient-based link scheduling optimization to reduce congestion and radio footprint.
\end{itemize}

\noindent
{\bf Notation:} 
$ |\cdot| $ and $\odot$ represent the cardinality of a set and Hadamard (element-wise) product operator. 
$ \mathbbm{1}(\cdot) $ is the indicator function.
$ \mathbb{E}(\cdot) $ stands for expectation. 
Upright bold lower-case symbol, e.g., $\bbz$, denotes a column vector, and we use $\bbz_i$ or $z_i$ denotes its $i$-th element. 
Upright bold upper-case symbol, e.g., $\bbZ$, denotes a matrix,  
and $\bbZ_{i,j}$ its element at row $i$ and column $j$, and $\bbZ_{*,j}$ for its column $j$ vector.
Calligraphic upper-case symbol denotes a set, e.g., $\ccalG$ for a graph,
and $ \ccalN_{\ccalG}(i) $ the set of immediate neighbors of node $i$ on graph $\ccalG$.

% \vspace{-0.1in}
\section{System Model and Problem Formulation}
\label{sec:problem}
% \vspace{-0.05in}
We model a wireless multi-hop network as a \emph{connectivity graph} $\ccalG^{n}$ and a \emph{conflict graph}~$\ccalG^{c}$. 
The connectivity graph is a {directed graph $\ccalG^{n}=(\ccalV, \ccalE)$, in which a node $i\in\ccalV$ denotes a device $i$, and a directed edge $(i,j)\in\ccalE$ represents that node $i$ can transmit data to node $j$ directly over-the-air.
Graph $\ccalG^{n}$ is assumed to be strongly connected, meaning that a directed path always exists between any pair of nodes.}
The conflict graph, $\ccalG^c=(\ccalE,\ccalH)$, describes the conflict relationship between links and is defined as follows: each vertex $e\in\ccalE$ corresponds to a link in $\ccalG^{n}$ and each undirected edge $(e_1, e_2)\in\ccalH$ indicates a conflict between links $e_1, e_2\in\ccalE$ in $\ccalG^{n}$. 
Two links are in conflict if they share the same transceiver (\emph{interface conflict}) or located closely such that their simultaneous transmissions can cause outage (\emph{wireless interference}). 
We assume that the conflict graph $\ccalG^c$ is known, 
e.g., obtained by local channel monitoring~\cite{zhao2022twc} or through advanced network-wide estimation techniques~\cite{yang2016learning}.

We consider wireless multi-hop networks with orthogonal multiple access (OMA) in a time-slotted MAC system.
The real-time link rate, $\acute\bbr_{e}(t) \in \mathbb{Z}^+$, is the number of packets that can be transmitted across link $e \in \mathcal{E}$ during time slot $t$ under fading channels. 
Vector $\mathbf{r} = [r_e \mid e \in \mathcal{E}]\in\reals^{|\ccalE|}$ collects the long-term link rates in the network, where $r_e = \mathbb{E}_t[\acute\bbr_{e}(t)]$ is the mean of $\acute\bbr_e(t)$ across time.

We define set $\ccalF$ to capture all the flows in the network, where a flow $f\in\ccalF$ could be a source-destination pair or a commodity that captures all the packets with the same destination.
We use matrix $\bbA\in\reals^{|\ccalV|\times|\ccalF|}$ to capture the exogenous packet arrival rates, where entry $\bbA_{i,f}$ denotes the exogenous packet arrival rate of flow $f$ on node $i$, where $\bbA_{i,f}>0$ if $i$ is a net source, and $\bbA_{i,f}<0$ if $i$ is a net sink. 
The routes of all flows are captured by a matrix $\mathbf{\Lambda} \in \reals_+^{|\mathcal{E}| \times |\mathcal{F}|}$, where entry $\bbLambda_{e,f}\geq 0$ represents rate assignment of flow $f$ on link $e \in \mathcal{E}$.
The routing matrix $\bbLambda$ is constrained by flow conservation, e.g., $ \bbDelta\bbLambda_{*,f} = \bbA_{*,f} $ for all $f\in\ccalF$, where $\bbDelta\in \reals^{|\ccalV| \times |\ccalE|}$ is the node-edge incidence matrix, in which a entry is defined as 
\begin{equation*}
     \bbDelta_{i,e} \doteq
    \begin{cases}
        +1, & \text{if link } e \text{ leaves node } i,\\[0.15em]
        -1, & \text{if link } e \text{ enters node } i,\\[0.15em]
        0,  & \text{otherwise}.
    \end{cases}
\end{equation*}
The assigned arrival rate on link $e \in \mathcal{E}$ is $\lambda_e = \sum_{f\in\mathcal{F}} \bbLambda_{e,f}$.

\begin{algorithm}[t!]
\caption{Modified Luby's Maximal Independent Set}
\label{algo:lubys}
\begin{algorithmic}[1]

\Require $\mathcal{G}^c$, 
         $\mathbf{z}\in\mathbb{R}^{|\mathcal{E}|}$,
         $\hat{\mathbf{b}}\in\{0,1\}^{|\mathcal{E}|}$, $M\ge1$

\Ensure Schedule $\mathbf{s}\in\{0,1\}^{|\mathcal{E}|}$

\State $\mathbf{s} = -\hat{\mathbf{b}}$,
       $\hat{\mathbf{b}}^{(1)} = \hat{\mathbf{b}}$,
       $m=1$

\While{$-1\in\mathbf{s}$ \textbf{and} $m\le M$}
    \State $\mathbf{c}\sim\mathbb{U}(\mathbf{0},\mathbf{z})$, \quad $\hat{\mathbf{c}}^{(m)} = \hat{\mathbf{b}}^{(m)} \odot \mathbf{c}$, \quad $\hat{\mathbf{b}}^{(m+1)} = \hat{\mathbf{b}}^{(m)}$

    \ForAll{link $e\in\ccalE$ with $\hat{\mathbf{b}}^{(m)}_e = 1$}
        \State Exchange $ \hat{\mathbf{c}}^{(m)}_e $ with neighbors $i\in\mathcal{N}(e)$
    
        \If{$\hat{\mathbf{c}}^{(m)}_e \!>\! \max\{ \hat{\mathbf{c}}^{(m)}_i \mid {i\in\mathcal{N}_{\ccalG^c}(e)}\}$}
            \State $\mathbf{s}_e = 1$, \ $\hat{\mathbf{b}}^{(m+1)}_e = 0$ \Comment{link $e$ scheduled, quit}
            \State Broadcast mute message to neighbors $i\in\mathcal{N}_{\ccalG^c}(e)$
        \ElsIf{Receive mute message from a neighbor}
            \State $\mathbf{s}_e = 0$, \ $\hat{\mathbf{b}}^{(m+1)}_e = 0$ \Comment{link $e$ muted, quit}
        \EndIf
    \EndFor

    \State $m \gets m+1$
\EndWhile

\end{algorithmic}
\end{algorithm}

We consider distributed link scheduling based on randomized contention, which can be formulated as finding a MIS on the conflict graph $\ccalG^c$, since a feasible schedule should be a set of non-conflicting links (independent set on $\ccalG^c$).
We generalize randomized contention as a weighted Luby's MIS algorithm~\cite{luby1985simple}, as detailed in Algorithm~\ref{algo:lubys} and denoted as $\bbs(t)=f_{\text{Luby}}(\ccalG^c, \bbz,\hat\bbb(t), M)$.
Here, $\bbs(t)\in\{0,1\}^{|\ccalE|}$ is the schedule of time slot $t$.
Vector $\mathbf{z} \in \mathbb{R}^{|\mathcal{E}|}$ sets link priorities by modifying the winning probabilities of a link.
Vector $\hat{\mathbf{b}}(t) \in \{0, 1\}^{|\mathcal{E}|}$ indicates which links are contending.
$M$ limits the maximum rounds of contention, for example, weighted and unweighted CSMA schemes~\cite{ni2012qcsma} can be captured by $M=1$.
In each contention round $m$, each undecided link (e.g., $\hat{\mathbf{b}}_e^{(m)} = 1$) redraws a random number $\hat{\mathbf{c}}^{(m)}_e \sim \mathbb{U}({0},\mathbf{z}_e)$ and compares it with neighbors.

Let $\mathbf{x} = [x_e \mid e \in \mathcal{E}] \in [0,1]^{|\mathcal{E}|}$ denote the vector of long-term duty cycles, where $x_e \doteq \mathbb{E}_t(s_e(t))$ and $s_e(t)\in\{0,1\}$. 
We consider the link duty cycle $\bbx$ as a function of network parameters and optimizable routing and scheduling policies:
\begin{equation}\label{eq:sys}
    \bbx = f_{\text{sys}} \big(\mathcal{G}^c, \mathbf{r}, \mathbf{\Lambda}, \mathbf{z}, M\big).
\end{equation}
The function $f_{\text{sys}}$ is usually obtained through costly packet-level network simulations, real-world experiments, or trainable NDT~\cite{li2023milcom,li2024glance,zhao2024tmlcn}, which are all resource intensive. 
To break these limitations, our study can be formalized as the following problem:
\begin{Problem}
\label{prob:ndt}
Approximate $f_{\text{sys}}$ in~\eqref{eq:sys} with an efficient network digital twin that generalizes across diverse network topologies and traffic patterns without costly simulation, training, or trial-and-error.
\end{Problem}

\section{Digital Twin for Randomized Link Scheduling}
\label{sec:solution}

We take two steps to develop the NDT $f_{\text{NDT}}$: 
In Section~\ref{sec:model}, we derive an analytical model of the weighted Luby's algorithm.
In Section~\ref{sec:dt}, we develop an iterative NDT based on this model.

\subsection{Analytical Model of Weighted Luby's MIS Algorithm}
\label{sec:model}
We denote a closed-form expression of link duty cycles induced by the weighted Luby's MIS in Algorithm~\ref{algo:lubys} as $\hat{\mathbf{x}} = f_{\text{model}}(\mathcal{G}^c, \mathbf{z}, \mathbf{B}, M)$, where $\mathbf{B}$ is the contending probability matrix.
The duty cycle of link $e$ accumulates contributions from all $M$ contention rounds:
\begin{equation}
    \hat{x}_e = \sum_{m=1}^M b_e^{(m)} P_{e,\text{win}}^{(m)},
\end{equation}
where $b_e^{(m)}$ is the probability that link $e$ participates in contention round $m$, e.g., $\Prcs{\hat{\bbb}^{(m)}_e=1}$, and $P_{e,\text{win}}^{(m)}$ is the conditional probability that link $e$ wins round $m$ when it contends in this round. 

Link $e$ wins round $m$ if its random number $\hat{\bbc}_e^{(m)}$ exceeds all competing neighbors. 
The winning probability is computed as
\begin{subequations}
\begin{align}
P_{e,\text{win}}^{(m)} &= \Prc{e\; \text{wins in round } m \mid \hat{b}^{(m)}_e=1} \\
&= \frac{1}{\mathbf{z}_e} \int_0^{\mathbf{z}_e} \prod_{i \in \mathcal{N}_{\mathcal{G}^c}(e)} F_{i|e}^{(m)}(x) \, dx \\
&= \frac{1}{\mathbf{z}_e} \int_0^{\mathbf{z}_e} \exp\left\{\sum_{i \in \mathcal{N}_{\mathcal{G}^c}(e)} \ln\left[F_{i|e}^{(m)}(x)\right]\right\} dx,\label{E:win:c}
\end{align}
\end{subequations}
Here, $F_{i|e}^{(m)}(x)$ is the conditional CDF of the random draw of neighbor $i$, given that link $e$ is contending. $F_{i|e}^{(m)}(x)$ is given by
\begin{equation}
\begin{aligned}
F^{(m)}_{i|e}(x) &= \Prc{\hat{\bbc}^{(m)}_i\le x \mid \hat{b}^{(1)}_e=1} \\
& =
\begin{cases}
0, & x<0,\\[3pt]
\left(1-b^{(m)}_{i\mid e}\right) + b^{(m)}_{i\mid e}\,\dfrac{x}{\mathbf{z}_i}, & 0\le x\le \mathbf{z}_i,\\[10pt]
1, & x>\mathbf{z}_i~.
\end{cases}
\end{aligned}
\end{equation}
where $b_{i\mid e}^{(m)}$ denotes the conditional probability that neighbor link $i$ participates in round $m$ given that $e$ does.

Conditional on link $e$ participating round $m$, i.e., $\hat{\mathbf{b}}_e^{(m)}=1$, the random number $\hat{\mathbf{c}}_e^{(m)}$ is uniformly distributed over $[0, \mathbf{z}_e]$ with density $f_e^{(m)}(x) = \frac{1}{\mathbf{z}_e}\,\mathbbm{1}_{[0,\mathbf{z}_e]}(x)$.
For efficient computation, we approximate the integral in~\eqref{E:win:c} using the $L$-point discretization:
\begin{equation}
P_{e,\text{win}}^{(m)} \approx \frac{1}{L} \sum_{l=0}^{L-1} \exp\left\{\sum_{i \in \mathcal{N}_{\ccalG^c}(e)} \ln\left[F_{i|e}^{(m)}\left(\frac{l \cdot \mathbf{z}_e}{L}\right)\right]\right\}.
\end{equation}
Link $e$ enters contention round $m+1$ if it did not win in round $m$ and none of its conflicting neighbors won. Thus, under a locally tree-like approximation~\cite{dembo2013factor}, the survival probability evolves  as 
\begin{equation}
b_e^{(m+1)} \approx b_e^{(m)} \left(1 - P_{e,\text{win}}^{(m)}\right) \prod_{i \in \mathcal{N}_{\mathcal{G}^c}(e)} \left(1 - b_{i|e}^{(m)} P_{i,\text{win}}^{(m)}\right).
\label{eq:approximation}
\end{equation}
Deriving from the queueing system, the initial contention probabilities for round 1 are given by
\begin{equation}
b_{e}^{(1)} = b^{in}_{e} = \min\left[\frac{\lambda_e}{\mu_e},1\right] , \quad b_{i\mid e}^{(1)} = \frac{b^{in}_{i,e}}{b^{in}_{e}}.
\end{equation}
where $\mu_e$ represents the effective capacity and is given by $\mu_e = x_er_e$.
The diagonal entry $b^{in}_e = b^{in}_{e,e}$ of $\mathbf{B}$ represents the marginal probability of link $e$ has packets to transmit, while $b^{in}_{i,e}$ represents the joint probability of links $i$ and $e$ are contending simultaneously. 

\begin{algorithm}[t]
\caption{Analytical Network Digital Twin for Luby's Scheduler} 
\label{algo:dt}
\begin{algorithmic}[1]

\Require $\mathcal{G}^c$, 
         $\mathbf{z}, \mathbf{r}\in\mathbb{R}^{|\mathcal{E}|}$, 
         $\Lambda\in\mathbb{R}^{|\mathcal{E}|\times|\mathcal{F}|}$,
         $M\ge 1$, $K\ge 1$, $\alpha\in(0,1]$
\Ensure $\widehat{\mathbf{x}}\in[0,1]^{|\mathcal{E}|}$

\State ${x}^{(0)}_e 
    = \dfrac{z_e}{\,z_e + \sum_{i\in\mathcal{N}(e)} z_i\,},\quad
       \lambda_e = \sum_{f\in\mathcal{F}} \Lambda_{e,f},
       \ \forall e\in\mathcal{E}$

\For{$k = 1,\dots,K$}

    % \State \textit{Update per-link contention probability}
    \State $\mu^{(k-1)}_e = r_e {x}^{(k-1)}_e, \forall e\in\mathcal{E}$ \Comment{link capacity estimation}
    \State $ b^{(k)}_e = \min\!\left[\frac{\lambda_e}{\mu^{(k-1)}_e},\,1\right],
           \ \forall e\in\mathcal{E}$ \Comment{Marginal cont. prob}

    % \State \textit{Update joint contention probability (independence assumption)}
    \State $b^{(k)}_{i,e} = b^{(k)}_i b^{(k)}_e,
           \ \forall (i,e) \in \ccalH$ \Comment{Joint prob w/ indep. asm}

    \State ${\dot{\bbx}}
        = f_{\text{model}}\!\left(
            \mathcal{G}^c,\,
            \mathbf{z},\,
            \bbB,\,
            M
          \right)$

    \State ${\mathbf{x}}^{(k)} 
        = \min\bigl\{(1-\alpha){\mathbf{x}}^{(k-1)} + \alpha\,{\dot{\bbx}},\;\; \mathbf{1}\bigr\}$

\EndFor

\State $\widehat{\mathbf{x}} \gets {\mathbf{x}}^{(K)}$
\end{algorithmic}
\end{algorithm}

\subsection{Iterative Algorithm for Network Digital Twin}
\label{sec:dt}
The analytical model $f_{\text{model}}$ in Section~\ref{sec:model} cannot be directly used as NDT since its input contending probability matrix $\bbB$ that is unavailable apriori. 
Therefore, we embed $f_{\text{model}}$ into  an iterative procedure $f_{\text{NDT}}$ that jointly refines duty cycles and contending probabilities~\cite{zhao2024congestionaware}, which shares the same input and output spaces as $f_{\text{sys}}$:
\begin{equation}
    \hat\bbx = f_{\text{NDT}} \big(\mathcal{G}^c, \mathbf{r}, \mathbf{\Lambda}, \mathbf{z}, M; K, \alpha\big).
\end{equation}
The detailed procedure is presented in Algorithm~\ref{algo:dt}.

The iterative NDT in Algorithm~\ref{algo:dt} exploits the circular dependency of variables in the analytical model, i.e., $\hat{\bbx}\rightarrow \bbmu \rightarrow \bbb \rightarrow \hat{\bbx}$. 
It initializes the $\hat\bbx$ with a conservative estimation $\bbx^{(0)}$ by assuming every link contends all the time (line 1).
In each iteration, e.g., $k$, the link capacity $\mu_e^{(k-1)}$ is updated based on $x_e^{(k-1)}$ (line 3), followed by updates of marginal and joint contending probabilities in lines 4-5. 
Next, the link duty cycle ${\bbx}^{(k)}$ is updated as a clipped exponential moving average (EMA) (line 7) of the estimates of analytical model $f_{\text{model}}$ (line 6).
The EMA ensures smooth evolution of ${\bbx}^{(k)}$.
Lastly, ${\bbx}^{(K)}$ is returned as the final estimation.

To simplify the computation, we employ the independence assumption (line 5)  due to the dominant role of marginal probability (see Fig.~\ref{fig:single:perlink}).
The convergence of the algorithm is ensured by its self-regulating property: smaller $x_e^{(k-1)}$ causes link $e$ to contend more often (larger $b_e^{(k)}$), which increases $x_e^{(k)}$ through the analytical model.
$K$ and $\alpha$ are given as parameters.
From experiments, we found that Algorithm \ref{algo:dt} converges quickly and $K=5$ is sufficient with $\alpha=0.5$.

\subsection{Complexity Analysis}
Algorithm~\ref{algo:dt} performs $K$ outer iterations of executing the analytical model, each has $M$ inner iterations. 
In each inner iteration, each link aggregates CDFs from its conflicting neighbors with a resolution of $L$, requiring $2|\ccalH|$ operations. 
The total computational complexity is $\mathcal{O}(K M L |\ccalH|)$.
Notice that the NDT can also be implemented in a fully distributed manner, of which
the complexity is measured as $\ccalO(MK)$ rounds of message exchanges, with a message size of $\ccalO(L)$.

\begin{figure*}[t]
\centering
 %\vspace{-0.1in}
% \hspace{-1.0mm}
\hspace{-2mm}
\subfloat[]{
    \includegraphics[height=1.74in]{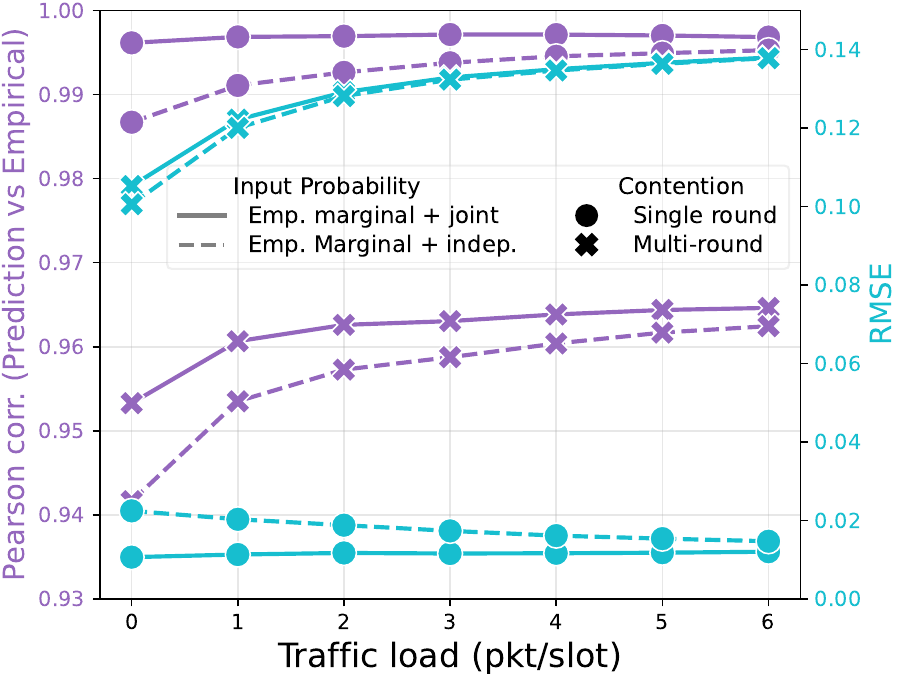}
    \label{fig:results:pearson}\vspace{-0.1in}
}
\hspace{-2mm}
\subfloat[]{
    \includegraphics[height=1.74in]{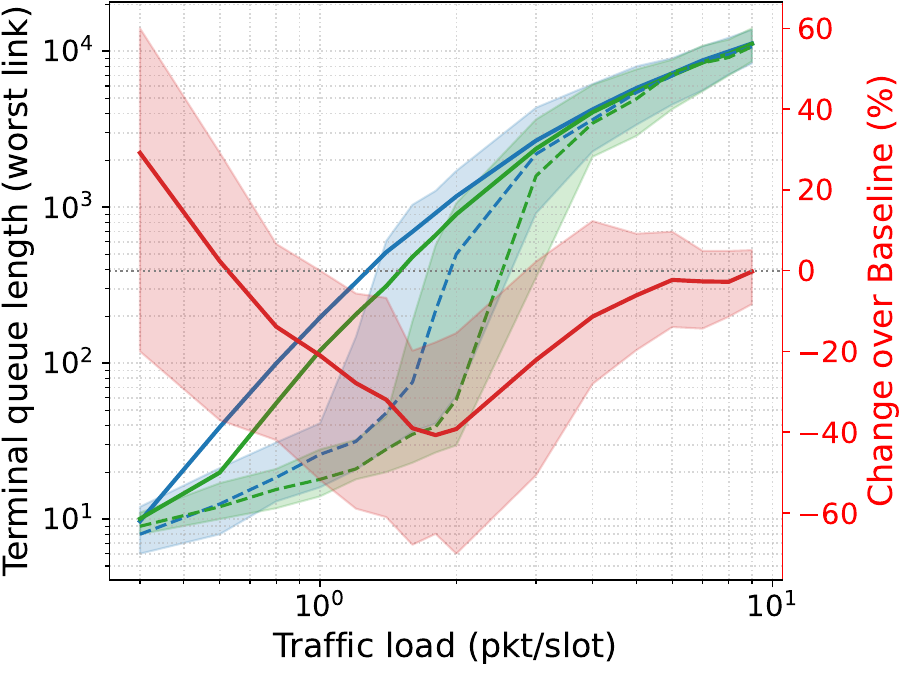}
    \label{fig:results:response}\vspace{-0.1in}
}
\hspace{-2mm}
\subfloat[]{
    \includegraphics[height=1.74in]{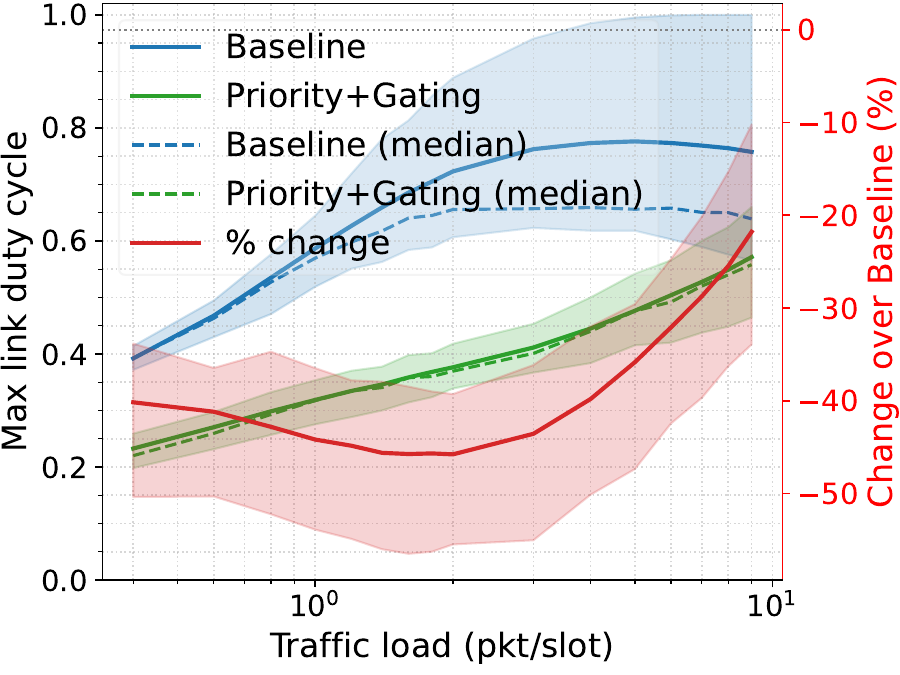}
    \label{fig:results:congest}\vspace{-0.1in}
}
    \vspace{-0.1in}
    \caption{Results of full-scale evaluations: (a)~Accuracy (Pearson correlation coefficient and root mean square error between predicted and empirical duty cycles) of the analytical model as a function of traffic load. This chart compares two input probability configurations and two contention types. 
    (b)~Maximum terminal queue length versus traffic load. 
    (c)~Per-instance maximum link duty cycle versus traffic load. 
    For (b) and (c), the secondary axis plots the percentage change relative to the Baseline, and the error bands stand for 25 and 75 percentiles respectively.
    }
 \label{fig:results}    
 \vspace{-0.2in}
\end{figure*}

% \vspace{-0.05in}
\section{Numerical experiments}
\label{sec:results}
% \vspace{-0.05in}
Our numerical evaluation is based on random wireless networks of sizes $|\mathcal{V}|\in\{20,50,100\}$ generated from a 2D point process, with nodes spread randomly and uniformly in a square area at a density of $8/\pi$. 
Links between two nodes are established if they are  within a distance of $1$. 
Interference is captured by the interface conflict model, assuming uniform transmit power and omnidirectional antennas. 
We test different traffic loads, $\beta\in[0.4,7]$,
for each pair of $(|\ccalV|,\beta)$, 100 test instances are generated from 10 different network topologies each with 10 realizations of random source-destination pairs, flow rates, and real-time link rates. 
For each instance, the number of flows is drawn from a discrete uniform distribution $|\ccalF|\in\mathbb{U}_{\mathbb{Z}}(\lfloor 0.15|\mathcal{V}| \rfloor, \lceil 0.25|\mathcal{V}| \rceil)$. 
For a flow $f\in\ccalF$, the exogenous packet arrival at its source follows a Poisson process with a constant rate of $\beta \cdot a_f$, where $a_f\sim\mathbb{U}(0.5, 1.5)$. 
The routing matrix $\mathbf{\Lambda}$ for each instance is found by shortest path routing and remains fixed during simulation. 
To capture fading channels with lognormal shadowing, the long-term link rate follows a uniform distribution $r_e\in \mathbb{U}(10, 42)$. The real-time rate $\acute{\bbr}_{e}(t)$ is subsequently modeled using a normal distribution $\mathbb{N}(r_e, 3)$. Each test instance is simulated over a horizon of $T = 1000$ time slots.

\begin{figure}[t]
    \centering 
    \subfloat[]{
    \includegraphics[height=1.5in]{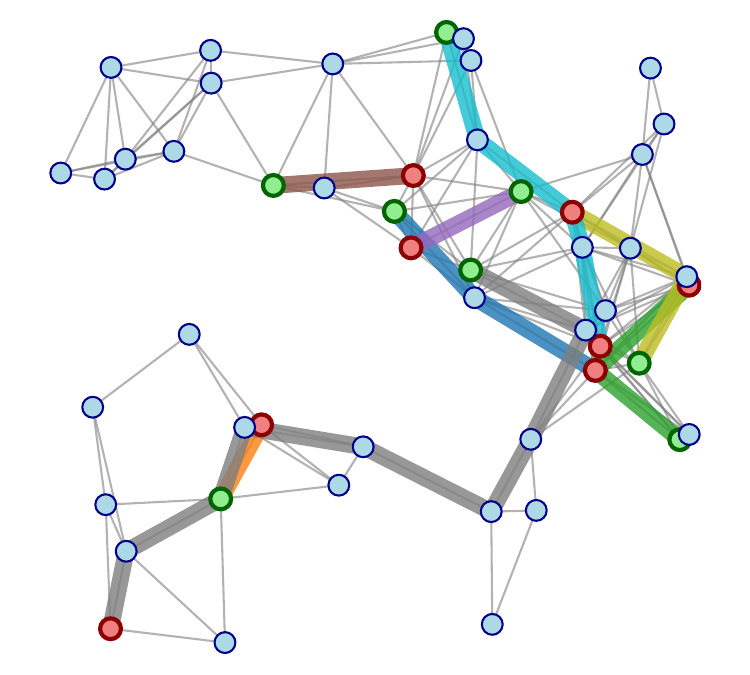}
    \label{fig:single:instance}\vspace{-0.1in}
}\hspace{-2mm}
\subfloat[]{
    \includegraphics[height=1.5in]{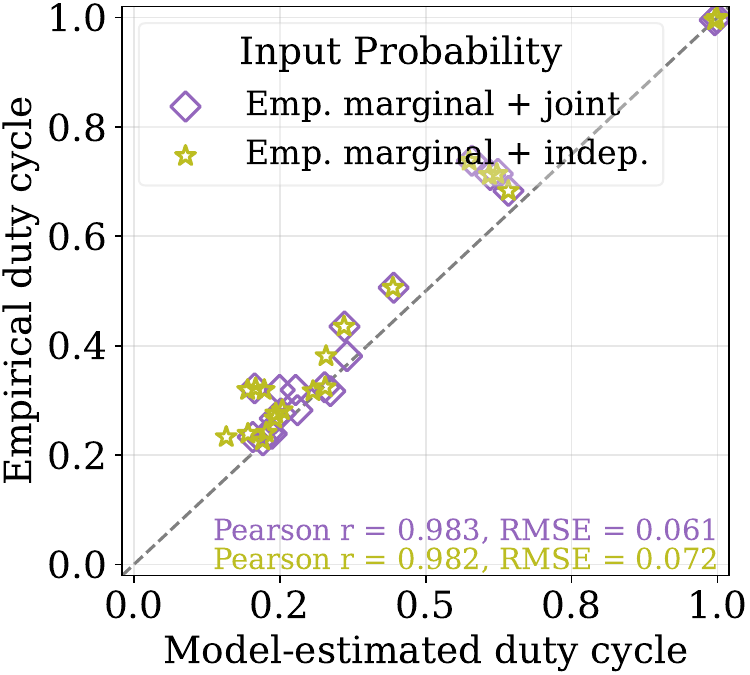}
    \label{fig:single:perlink}\vspace{-0.1in}
}\\
\subfloat[]{
    \includegraphics[height=1.5in]{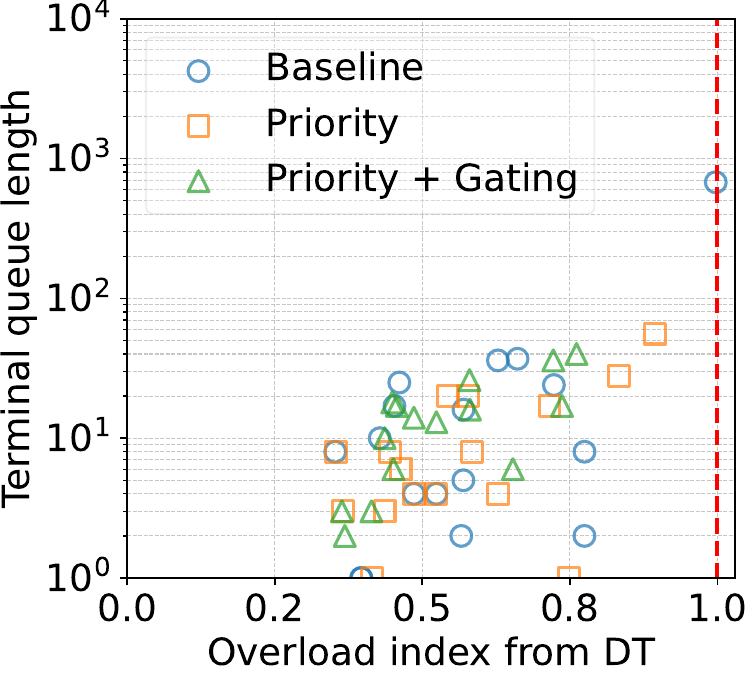}
    \label{fig:single:queue_length}\vspace{-0.1in}
}
\subfloat[]{
    \includegraphics[height=1.5in]{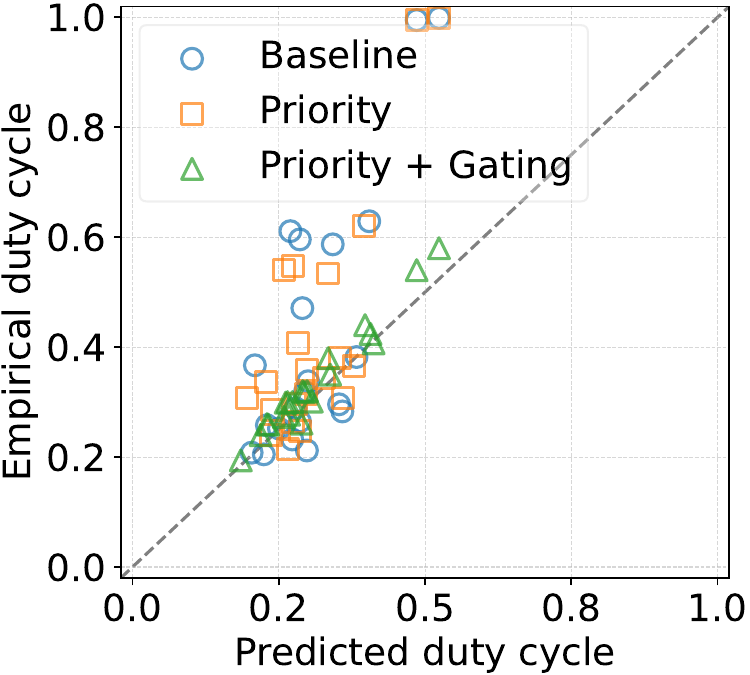}
    \label{fig:single:duty_cycle}\vspace{-0.1in}
}
\vspace{-0.05in}
\caption{Per-link results on a single test instance: 
(a) network topology and flow routes (green: source, red: destination, thick edges: route), 
(b) Empirical vs model-estimated link duty cycles with different input contention probabilities, 
(c) terminal queue length vs overload index from NDT under tested scheduling strategies,
and 
(d) Empirical link duty cycle vs NDT prediction under tested scheduling strategies. 
}
\label{fig:single}
\vspace{-0.1in}
\end{figure}

We demonstrate the accuracy of our analytical model and the gradient-based link scheduling optimization enabled by our NDT.
The results are presented 
in Figs.~\ref{fig:results} for the full-scale tests,
and in Figs.~\ref{fig:single} for detailed per-link comparisons under a single instance.  

\subsection{Analytical Model Accuracy}

We evaluate the accuracy of analytical model by comparing its predicted link duty cycles against the simulated empirical values.
Two configurations of the input probability matrix $\mathbf{B}$ are compared: (1) marginal probabilities $b_e^{in}$ only, assuming link independence, i.e., $b_{i,e}^{in} = b_e^{in} \cdot b_i^{in}$, and (2) both marginal and joint probabilities $b_{i,e}^{in}$. 
These inputs are empirical probabilities obtained from preliminary simulations. 
To illustrate per-link performance, we present the results of a test instance with 50 nodes, 8 flows, and $\beta=5.0$ as shown in Fig.~\ref{fig:single:instance}.
Fig.~\ref{fig:single:perlink} demonstrates a close correspondence between predicted and empirical duty cycles, with Pearson correlations exceeding $0.98$ and RMSE around $0.06\sim0.07$.
Fig.~\ref{fig:results:pearson} illustrates the prediction accuracy as a function of traffic load. 
Single-round contention achieves Pearson correlations above $0.99$ with RMSE below $0.02$ across all load levels. 
In contrast, multi-round contention based analytical model exhibits larger prediction errors ($0.10\sim0.14$) that increase by load due to the approximations in \eqref{eq:approximation}.

\subsection{Applications in Link Scheduling Optimization}

Next, we demonstrate the applications of our NDT in gradient-based link scheduling optimization, where the objective is to reduce congestion and energy consumption in the network under a prescribed routing scheme $\bbLambda$.
We optimize link priority vector $\bbz$ by using Adam optimizer to minimize the following loss function: 
$$
\ell(\bbz) = \frac{1}{|\ccalE|}\sum_{e\in\ccalE} \sigma(3(\rho_e-0.8)) + \max\{\rho_e-1,0\}\;, \quad \rho_e=\frac{\lambda_e}{\hat{x}_e r_e}\;,
$$
where $\sigma$ is sigmoid function, and $\rho_e$ is the overload index predicted by NDT, as $\hat{\bbx}=f_{\text{NDT}}(\ccalG^c,\bbr,\bbLambda,\bbz,M;\; K,\alpha)$. 
We evaluate the following three link scheduling policies:
a) \textit{Baseline}: all links have equal priority $\bbz=\boldsymbol{1}$.
b) \textit{Priority}: using link priority vector $\tilde{\bbz}$ optimized by 20 steps of gradient descent. 
c) \textit{Priority + Gating}: Besides adopting link priority $\tilde{\bbz}$, it further lets a link $e$ skip the contention at $t$ if its empirical link duty cycle in a sliding window of 100 time slots exceeds $1.1\tilde{x}_e$, where $\tilde{\bbx} = f_{\text{NDT}}(\ccalG^c,\bbr,\bbLambda,\tilde\bbz,M;\; K,\alpha)$ is the predicted link duty cycle.

\subsubsection{Runtime Comparison}

\begin{table}[t]
\centering
\caption{Average Runtime: Analytical NDT versus Simulation\protect\footnotemark}\vspace{-1mm}
\label{tab:runtime_compact}
\footnotesize
\begin{tabular}{@{}c|c|c|c|c@{}}
\hline\hline
\textbf{Network Size} & \textbf{Load} & \textbf{Analytical NDT (s)} & \textbf{Simulation (s)} & \textbf{Speedup} \\
\hline
\multirow{2}{*}{20}
   & 1.0 & 0.00272 & 1.718  & 631.5 \\
   & 7.0 & 0.00298 & 3.563  & 1195.6 \\
\hline
\multirow{2}{*}{50}
   & 1.0 & 0.00517 & 8.368  & 1618.2 \\
   & 7.0 & 0.00534 & 14.382 & 2695.3 \\
\hline
\multirow{2}{*}{100}
   & 1.0 & 0.00832 & 28.726 & 3450.9 \\
   & 7.0 & 0.00821 & 44.545 & 5424.0 \\
\hline\hline
\end{tabular}
\end{table}
\footnotetext{Experiments were conducted on a MacBook Pro with Apple M1 Pro chip (10-core CPU, 16-core GPU) and 16GB RAM.}

Table \ref{tab:runtime_compact} summarizes the average runtime of the NDT versus our efficient in-house packet-level simulation\footnote{Code available: \url{https://github.com/JoieMing/luby-ndt/}}. 
NDT computes steady-state duty cycles in milliseconds, achieving a speedup of three orders of magnitude for 100-node networks compared to simulation. 
While simulation runtime scales with both network size and traffic load, NDT requires only a single execution to predict mean-field results, scaling primarily with network size. 
For a 100-node network, NDT consistently finishes in $\approx 8$ ms, whereas simulation requires $28$ to $44$ seconds depending on traffic load.

\subsubsection{Congestion Mitigation}
Figure~\ref{fig:single:queue_length} illustrates per-link terminal queue lengths in the selected test instance. Priority reduces the congestion on the worst link compared to baseline by optimizing link priorities using the NDT. 
Priority+Gating further reduce the worst terminal backlog with restrained contention guided by NDT.

Figure~\ref{fig:results:response} presents the worst terminal queue length as a function of traffic load. 
Under load $ \beta=1 $, the median terminal queue length of the worst link is reduced by one order of magnitude compared to baseline.  
The gating mechanism suppresses links when their empirical duty cycle exceeds the predicted value, increasing the winning probabilities of neighboring links. 
Under heavier loads, congestion persists mainly due to greedy shortest path routing, where link priority optimization could offer little help.

\subsubsection{Link Duty Cycle}
Figure~\ref{fig:single:duty_cycle} shows that empirical and predicted duty cycles are well aligned under Priority+Gating policy, but exhibit large gaps under baseline and Priority policies. 
This gap is not due to inaccurate NDT predictions, but rather unrestricted contention. 
Empirical duty cycles can exceed the need of actual traffic demands when links operate in less crowded neighborhoods, increasing radio footprint (interference) and energy consumption. 
This is further confirmed by the per-instance maximum link duty cycle as a function of traffic load under the full-scale test results, as illustrated in Fig.~\ref{fig:results:congest}.
In addition to reduced congestion as shown in Figure~\ref{fig:results:response}, Priority+Gating lowers the maximum duty cycles from the baseline across all traffic loads.
Under moderate loads ($\beta \in [1,4]$), the maximum duty cycle is reduced by up to $42\%$, as load increases, the reduction narrows due to bottlenecks that scheduling optimization alone cannot eliminate. 
In short, NDT-guided Priority+Gating policy simultaneously reduces resource wastage and congestion.

\section{Conclusions}
\label{sec:conclusions}
% \vspace{-0.05in}
We present an analytical digital twin for wireless multi-hop networks with randomized contention that accurately predicts link duty cycles and congestion while requiring only lightweight computation and minimal input data. 
The NDT delivers high accuracy, especially for single-round contention, and offers substantial speed advantages over packet-level simulation.
Beyond modeling, the NDT provides explicit gradients with respect to the tunable routing matrix and scheduling priorities, enabling the first gradient-based tuning of link-level priorities and opening a new direction for policy optimization in wireless networks.
Future directions include extending this gradient-based framework to min-cost flow and routing optimization under nonlinear wireless constraints, as well as developing lightweight real-time distributed implementations that can be embedded into next-generation MAC protocols for autonomous networking.

% -------------------------------------------------------------------------

\bibliographystyle{ieeetr}
\bibliography{strings}

\end{document}